\documentclass{article}
\usepackage{ijcai25}

\usepackage{times}
\usepackage{soul}
\usepackage{url}
\usepackage[hidelinks]{hyperref}
\usepackage[utf8]{inputenc}
\usepackage[small]{caption}
\usepackage{graphicx}
\usepackage{amsmath}
\usepackage{mathrsfs}
\usepackage{amsthm}
\usepackage{booktabs}
\usepackage{algorithm}
\usepackage{algorithmic}
\usepackage[switch]{lineno}
\usepackage{multirow}
\usepackage{amssymb}
\usepackage{xcolor}
\usepackage{bm}
\usepackage{svg}
\usepackage{float}

\makeatletter
\renewcommand{\maketag@@@}[1]{\hbox{\m@th\normalsize\normalfont#1}}%
\makeatother

\usepackage{xurl}

\definecolor{myred}{HTML}{ea3323}
\definecolor{myblue}{HTML}{4eadea}
\newcommand{\redtext}[1]{\textcolor{myred}{#1}}
\newcommand{\bluetext}[1]{\textcolor{myblue}{#1}}

\title{Progressive Uncertainty-Guided Evidential U-KAN for Trustworthy Medical Image Segmentation}

\author{
Zhen Yang
\and
 Yansong Ma\and
Lei Chen
}

\begin{document}

\maketitle

\begin{abstract}
Trustworthy medical image segmentation aims at deliver accurate and reliable results for clinical decision-making. Most existing methods adopt the evidence deep learning (EDL) paradigm due to its computational efficiency and theoretical robustness. However, the EDL-based methods often neglect leveraging uncertainty maps rich in attention cues to refine ambiguous boundary segmentation. To address this, we propose a progressive evidence uncertainty guided attention (PEUA) mechanism to guide the model to focus on the feature representation learning of hard regions. Unlike conventional approaches, PEUA progressively refines attention using uncertainty maps while employing low-rank learning to denoise attention weights, enhancing feature learning for challenging regions. Concurrently, standard EDL methods suppress evidence of incorrect class indiscriminately via Kullback-Leibler (KL) regularization, impairing the uncertainty assessment in ambiguous areas and consequently distorts the corresponding attention guidance. We thus introduce a semantic-preserving evidence learning (SAEL) strategy, integrating a semantic-smooth evidence generator and a fidelity-enhancing regularization term to retain critical semantics. Finally, by embedding PEUA and SAEL with the state-of-the-art U-KAN, we proposes Evidential U-KAN, a novel solution for trustworthy medical image segmentation. Extensive experiments on 4 datasets demonstrate superior accuracy and reliability over the competing methods. The code is available at \href{https://anonymous.4open.science/r/Evidence-U-KAN-BBE8}{github}.
\end{abstract}

\section{Introduction}
Medical image segmentation, as a core technology in computer-aided diagnosis, aims to precisely identify target regions such as organs and lesions through pixel-level classification, playing a crucial role in disease diagnosis and treatment planning. Early research primarily relied on traditional image processing techniques, such as region segmentation \cite{zhang2015medical}, clustering \cite{christ2011segmentation} 
\begin{figure}[H]
    \centering 
    \includegraphics[width=1\linewidth, angle=0]{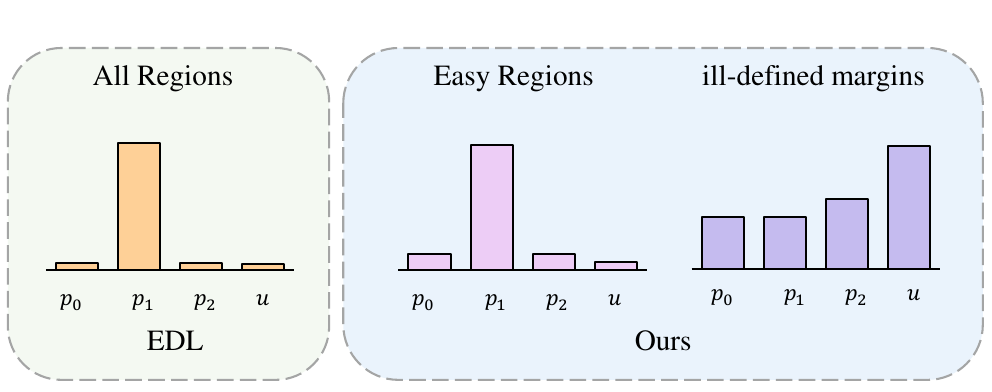}
    \caption{Compared to traditional evidential deep learning methods, Semantic-preserving evidence learning strategy is capable of generating more precise uncertainty estimates and evidence distributions in ill-defined margins. Not only does it achieve superior guidance effects through the progressive evidence uncertainty guided attention, but it also significantly enhances the accuracy and robustness of uncertainty quantification. Particularly in scenarios with ambiguous class boundaries and uneven sample distributions, our method demonstrates higher reliability and stability.}
    \label{fig:SAEL}
\end{figure}
\noindent 
and edge detection \cite{salman2015medical}, etc. Although these methods provided solutions to some extent, they struggled to balance accuracy and efficiency when dealing with complex medical images. 
While these methods offer a solution to some extent, it is often challenging to consider both segmentation accuracy and efficiency when dealing with complex medical images.\par
The breakthrough in deep learning technology has brought about a paradigm shift in medical image segmentation. The encoder-decoder architecture, represented by UNet proposed by Ronneberger et al.  \cite{ronneberger-etal:u-net} , has demonstrated significant advantages in medical image segmentation tasks by fusing multi-scale features through skip connections. Subsequent improved models, such as UNet++  \cite{zhou-etal:unet++}, UNet3+ \cite{huang-etal:unet3}, and nnU-Net \cite{isensee-etal:nnu-net}, have further enhanced the model's ability to capture subtle structures. However, these methods rely on the classification confidence output by Softmax as the basis for segmentation, which can lead to overconfident predictions for out-of-distribution (OOD) samples, such as rare lesion morphologies.\par
%
%
%
%
To address this issue, researchers have proposed evidence deep learning (EDL) based on the Dirichlet distribution \cite{sensoy-etal:evidential}.
EDL models cognitive uncertainty, enabling the model to estimate the reliability of predictions. However, due to the global penalty mechanism (KL divergence regularization) for incorrect categories during training \cite{sensoy-etal:evidential}, the model's ability to extract features in ambiguous or poorly defined edge regions is compromised \cite{bao-etal:evidential-open-set}, thereby affecting segmentation accuracy \cite{chen-etal:uncertainty-aware}. 
On the other hand, existing works only use uncertainty maps as posterior evaluation metrics, and the rich attention-guiding information contained in uncertainty maps has not been fully utilized. This information can improve segmentation accuracy and robustness by guiding the model's attention. \par
%
To tackle the aforementioned challenges, we propose a progressive evidence learning framework based on U-KAN\cite{li-etal:ukan}, as a network architecture with powerful function approximation capabilities, provides a solid foundation for our method. We design a progressive segmentation approach that effectively utilizes the attention-guiding information in uncertainty maps. This design uses uncertainty maps as key inputs, dynamically guiding the model to focus on challenging or ambiguous regions through cross-attention, thereby improving segmentation accuracy and robustness. To ensure the reliability of uncertainty guidance in poorly defined edge regions during the guidance process, we introduce an evidence generation function that favors semantic alignment and a corresponding evidence uncertainty regularization term. This ensures that semantic information of the target regions is fully considered during evidence generation, avoiding the evidence compression problem in traditional EDL methods at boundary regions. This enhances KAN's uncertainty estimation capability, better supporting the progressive evidence uncertainty-guided attention mechanism, and further improving the model's segmentation performance in ambiguous regions.
%
%
\par
In summary, the main contributions of this paper are as follows: 
\begin{itemize}
    \item We proposed a novel progressive evidence learning framework built on U-KAN, leveraging evidential uncertainty and attention-guided information from uncertainty maps to enhance segmentation.
    \item We devised a core design that uses uncertainty maps as input to steer segmentation via cross-attention, dynamically focusing on challenging or ambiguous regions to improve accuracy and robustness.
    \item We introduced an evidence generation function and  corresponding evidence uncertainty regularization term, avoiding the evidence compression problem in traditional EDL methods at boundary regions, and enhancing the model's feature extraction capability in ambiguous regions.
\end{itemize}


\begin{figure*}[ht]
    \centering
    \includegraphics[width=0.90\linewidth]{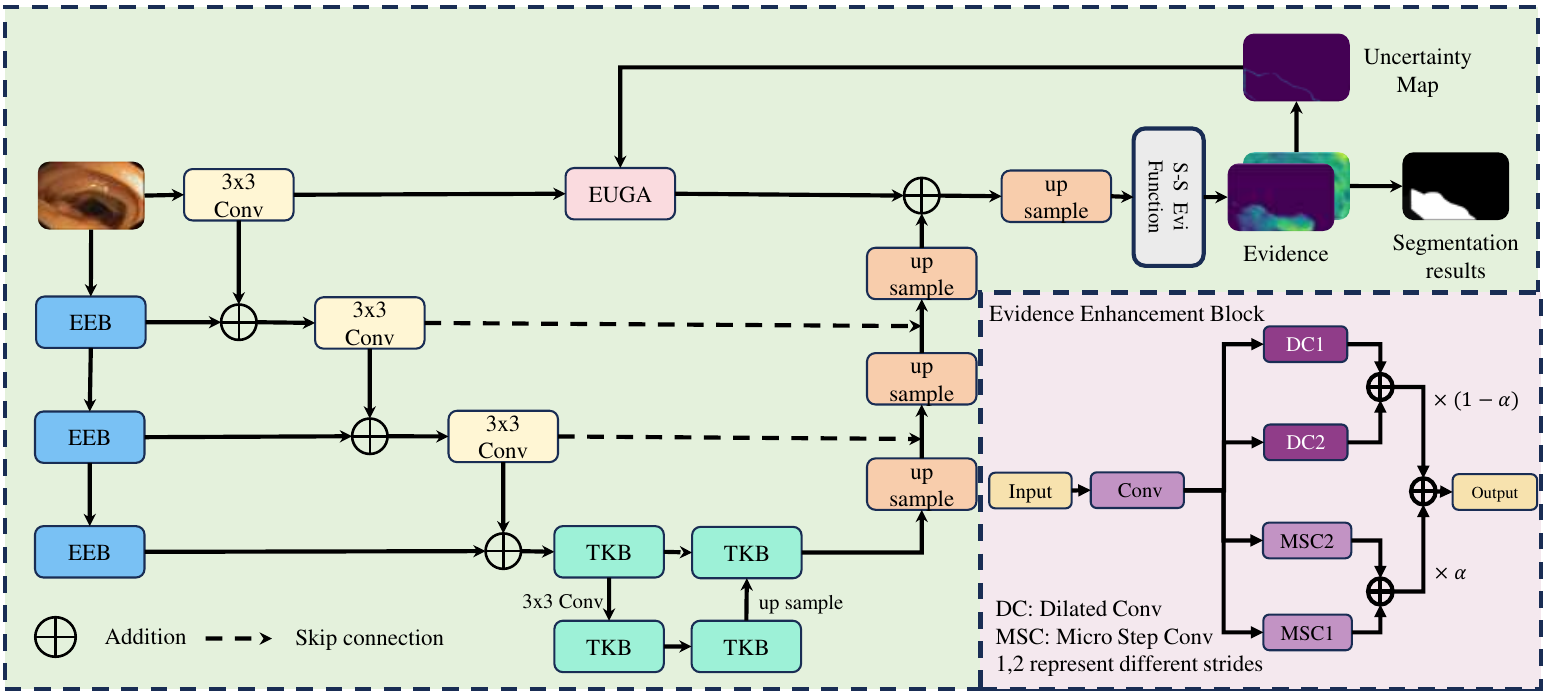}
    \caption{The overview of the proposed framework. Evidential U-KAN is an iterative network based on a U-KAN with five down-sampling stages. In the shallow layers, we propose the Evidence Enhancement Block (EEB) to integrate multi-scale information and enhance the model's capability for evidence extraction. In the deeper layers, the Tokenized KAN Block (TKB) replaces the MLP to enhance the model's learning capacity. The network replaces the skip connection of the shallowest layer with the Evidential Uncertainty Guided Attention module (EUGA), using the uncertainty map from the previous iteration (initialized as an all-ones matrix) as a condition to guide the model's segmentation. The iteration process stops once the convergence condition is satisfied.}
    \label{fig:enter-label}
\end{figure*}
\vspace{-8pt}

\begin{figure*}[ht]
    \centering
    \includegraphics[width=1\linewidth]{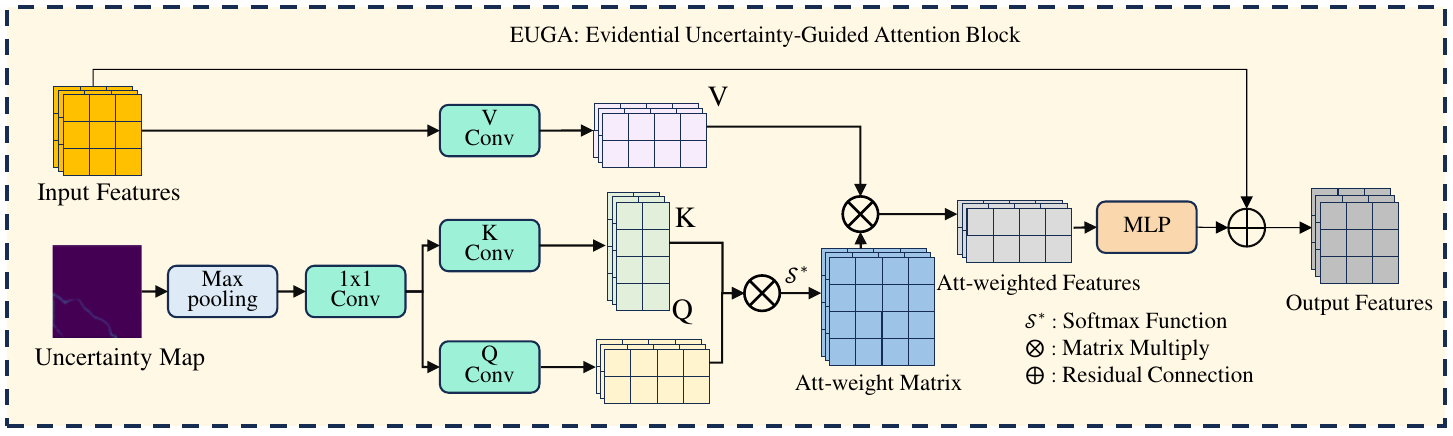}
    \caption{The overview of our Evidential Uncertainty Guided Attention module.We use a self-attention-like mechanism to extract an attention weight matrix from the uncertainty map. This matrix is then applied to weight the extracted input image features, enhancing key features and suppressing redundant information. To preserve the original input features, a residual connection is added at the end, combining the weighted features with the original input through element-wise addition.}
    \label{fig:EUGA_module}
\end{figure*}
\section{Related Work}
\subsection{Medical Image Segmentation}
Medical image segmentation is a technique for accurate segmentation of specific anatomical structures or pathological features of interest in medical images. With the introduction of UNet \cite{ronneberger-etal:u-net}, U-shaped neural network became the dominant paradigm for medical image segmentation modeling However, its fixed receptive field makes it challenging to model global context.  To address it, subsequent advancements include: UNet++\cite{zhou-etal:unet++}, UNet3+\cite{huang-etal:unet3} and nnU-Net\cite{isensee-etal:nnu-net}. Based on this, DC-UNet\cite{lou-etal:dc-unet} integrates multi-resolution convolution to extract features from images. DeepLabv3\cite{yurtkulu-etal:deepLabv3}, DeepLabv3+\cite{liu-etal:satellite-marsh} fuses multi-scale information by introducing spatial pyramids. Despite these improvements, CNN-based methods are limited by the inductive bias of local convolution operations, making it difficult to handle long-range dependencies in complex anatomical structures. In recent years, the introduction of the Transformer\cite{vaswani:attention} into medical image segmentation, particularly with the proposal of ViT\cite{dosovitskiy-etal:image-worth},, has brought breakthrough progress: TransUNet\cite{chen-etal:transunet} combines CNN and Transformer in a hybrid architecture to leverage global modeling capabilities, improving boundary segmentation accuracy. Swin-Unet\cite{cao-etal:swin-unet} reduces computational complexity through shifted windows. nnFormer\cite{zhou-etal:nnformer} further combines both CNN and Transformer heuristically to improve the segmentation accuracy of the model. Although these methods significantly improve global feature modeling capabilities, their large parameter count and high computational cost limit their clinical deployment potential. \par
Based on the Kolmogorov-Arnold representation theorem, KAN\cite{liu-etal:kan} bhas demonstrated function approximation capabilities surpassing those of MLP and has been applied across multiple fields.  U-KAN\cite{li-etal:ukan} marks the first integration of KAN with the UNet architecture, enhancing fine-grained feature extraction efficiency through learnable activation functions, thus endowing the network with stronger feature extraction capabilities.\par
\subsection{Uncertainty Estimation}
Traditional semantic segmentation models interpret the category probabilities output by the Softmax function as classification confidence, which often leads to the problem of “overconfidence”\cite{han-etal:multi-view}\cite{vanamersfoort-etal:uncertainty-estimation}. Such models exhibit high confidence even when the predictions are wrong. In order to address this issue, many uncertainty estimation methods have been introduced. Traditional uncertainty estimation methods mainly include Ensemble\cite{lakshminarayanan-etal:deep-ensembles}, Bayesian-based(Monte-Carlo Dropout)\cite{kendall-gal:bayesian-uncertainties} and Evidential Deep Learning\cite{sensoy-etal:evidential}. Among these, uncertainty estimation methods based on Evidential Deep Learning (EDL) are gaining more and more attention. The core of EDL is “evidence”, which represents the degree of belief or support for different assumptions in the model predictions based on Dempster-Shafer Theory (DST). As the model is trained, it learns to assign high evidence to correct predictions and low evidence to incorrect predictions. Evidential Deep Learning was initially used for classification task uncertainty estimation, \cite{sensoy-etal:evidential} pioneered the use of subjective logic theory along with DST to directly model classification uncertainty in deep neural networks, which significantly improved the robustness of neural networks to adversarial perturbations. \cite{zou-etal:tbrats} first introduced evidential learning to medical image segmentation tasks. \cite{chen-etal:uncertainty-aware} succeeded in introducing an evidential uncertainty-aware approach into a semi-supervised medical image segmentation task. \cite{yang-etal:lung-nodule} combined multiple evidences to guide segmentation through evidence fusion. Existing studies [23-25] typically use uncertainty maps as posterior evaluation metrics but have not explored their potential as prior guiding information. The \textbf{Progressive Evidential Uncertainty-guided Attention (PEUA)} proposed in this paper, through iterative optimization and low-rank decomposition, achieves dynamic and adaptive uncertainty attention mapping for the first time, addressing a methodological gap in this field.\par
\section{Methodology}
As demonstrated in Figure \ref{fig:enter-label},  the overall framework of our Evidential U-KAN, involving an encoder-decoder structure and incorporates evidential learning theory. It consists of three key components: the Evidential Uncertainty Guided Attention (EUGA) module, evidence generation functions that facilitate semantic alignment, and the corresponding evidence uncertainty regularization term. Specifically, for the input image \textit{I}, it is first passed through the encoder model (U-KAN with enhanced multi-scale feature extraction capability) to extract multi-level features. The EUGA module, introduced in Section \ref{sec:PEUGA}, is then used to extract the feature information from the uncertainty map and interact it with the shallowest layer features using an attention mechanism to obtain the uncertainty-guided feature map. After performing skip connections with the corresponding layers of the decoder, the obtained \textit{logit} are processed by the evidence generation function introduced in Section 3.2 to produce the evidence output and calculate the corresponding prediction and uncertainty map. This process is iterated multiple times, with the iterations stopping when the difference in the uncertainty maps between adjacent iterations becomes smaller than a threshold, resulting in the final segmentation prediction and the corresponding uncertainty map.\par
\subsection{Progressive Evidence Uncertainty guided Attention}
\label{sec:PEUGA}
\subsubsection{Evidential Uncertainty Guided Attention }
In traditional evidential deep learning, uncertainty map is typically provided to clinicians only as part of the prediction results to assess the reliability of segmentation outcomes. However, uncertainty map actually contains rich attention information, which can effectively identify regions in the image where the model struggles to achieve accurate segmentation. These regions often include complex anatomical structures or noise interference, making them critical areas for the model to focus on. Based on this insight, we designed the EUGA module, aiming to leverage the attention information in uncertainty map to guide the model to focus more on high-uncertainty regions, thereby improving segmentation performance.\par
As demonstrated in Figure \ref{fig:EUGA_module}, to extract the main structural information from the uncertainty map as guidance while filtering out high-frequency, low-uncertainty noise, we introduce the concept of low-rank decomposition. Low-rank decomposition approximates the original data with a simpler, lower-dimensional matrix, preserving its main structural information while removing redundancy and noise. Specifically, we process the uncertainty map through two distinct convolutional operations to obtain two feature matrices, K and Q. These matrices have a significantly smaller dimension compared to the uncertainty map. The attention weight matrix is then derived through matrix multiplication and the softmax function. This operation not only reduces computational complexity but also retains the key region information in the uncertainty map relevant to the segmentation task. Additionally, to avoid disrupting the original data flow and to fully utilize the information in the input feature map, we incorporate a residual connection mechanism. This mechanism fuses the original feature map with the attention-weighted feature map. Such a design introduces attention-guided information while preserving the spatial details and semantic information of the original features, further enhancing the model's robustness and segmentation accuracy.\par
\subsubsection{Progressive Evidential Uncertainty Guided Segmentation Algorithm}
To maximize the attention-guiding role of uncertainty maps, we propose a Progressive Evidential Uncertainty Guided Segmentation Algorithm. The core idea of this algorithm is to iteratively optimize the uncertainty map, gradually guiding the model to focus on regions with high uncertainty, thereby improving segmentation accuracy. Specifically, for each input image, the model initializes an all-ones matrix as the initial uncertainty map, indicating that the model has maximum uncertainty for all regions at the initial stage. After the \textit{i}-th iteration, the model updates the uncertainty map based on the current output and uses it as the guiding information for the (\textit{i + 1})-th iteration. This process continues until the uncertainty map $u_i$ satisfies the convergence criteria as follow:

\begin{equation}
    Average(|u_{i+1}-u_i |) \leq \varepsilon
\end{equation}

The pseudocode of the algorithm is demonstrated in Algorithm \ref{alg:algorithm}. In the algorithm, we define the following notations: Let $f(\cdot)$ denote the model, let $\bm{J_{m,n}}$ denote the $m \times n$ all-one matrix, let $\bm{e_i}$ denote the evidence collected by the model, let $\bm{\alpha_i}$ denote the parameter of the Dirichlet Distribution,  let $\bm{S_i}$ denote the Dirichlet strength, let $\bm{u_i}$ denote the segmentation uncertainty, let $\bm{U}$ denote the Uncertainty Map matrix recorded inside the model, let $\bm{p_i}$ denote the expected prediction probability, and let $\bm{P}$ denote the prediction result.\par
\begin{algorithm}[htbp]
    \caption{Progressive Evidential Uncertainty Guided Segmentation Algorithm}
    \label{alg:algorithm}
    \textbf{Input}: Medical image $\bm{X_i} \in \mathbb{R}^{3\times m \times n}$ \\
    \phantom{\textbf{Input}: }Uncertainty map $\bm{U} \in \mathbb{R}^{3\times m \times n}$\\
    \textbf{Output}: Prediction mask $\bm{P} \in \mathbb{R}^{m \times n}$ \\
    \phantom{\textbf{Output}: }$\bm{U} \in \mathbb{R}^{3\times m \times n}$
    \begin{algorithmic}[1] 
        \STATE $\bm{U} \leftarrow \bm{J_{m,n}}$
        \WHILE{$\frac{1}{m \times n} || \bm{U_i} - \bm{U} ||_1 > \epsilon$}
        \STATE $\bm{e_i} = f(\bm{X_i})$
        \STATE $\bm{\alpha_i} = \bm{e_i} + 1$
        \STATE $\bm{S_i} = \sum_c^C\bm{\alpha_{ic}}$
        \STATE $\bm{U_i} = C/\bm{S_i}$
        \STATE $\bm{U} \leftarrow \bm{U_i}$
        \ENDWHILE
        \STATE $\bm{p_i} = \bm{\alpha_i}/\bm{S_i}$
        \STATE $\bm{P} = \underset{p_i}{\mathrm{arg\,max}} \, \bm{p_i}$
        \RETURN $\bm{P}, \bm{U}$
    \end{algorithmic}
\end{algorithm}

\subsection{Semantic-Preserving Evidence Learning Strategy}

\subsubsection{The Theory of Evidence for Medical Image Segmentation}
Unlike traditional medical image segmentation methods, evidential deep learning models the output of the network as a Dirichlet distribution, which is considered the conjugate prior of the multinomial distribution. This modelling approach not only provides a predictive distribution for medical image segmentation but also quantifies the uncertainty from this distribution, thus offering theoretical support for evaluating the reliability of segmentation results. Specifically, for a C-category input  pixel $x$, the evidence vector $\bm{e}_x$ corresponding to the network output can be used to calculate the Dirichlet distribution parameters $\bm{\alpha}_x$  and $S$ through the following function:
\begin{equation}
    \alpha^c_x=e^c_x+1,\ \ S=\sum_{c=1}^{C}\alpha_x^c
\end{equation}

These parameters essentially describe the probability distribution of each pixel belonging to different categories, where the expected probability and uncertainty for category c can be calculated using the following formula:
\begin{equation}
    \hat{p}_x^c = \frac{\alpha_x^c}{S}, u_x = \frac{C}{S}
\end{equation}

\subsubsection{Fidelity-Enhancing Regularization Term}
For a C-category classification task, the likelihood loss function can be written in the following form:

\begin{equation}
    \mathcal{L}_{CE}\left(p_x^C,y_x^C\right)=\prod_{c=1}^{C}\left(p_x^c\right)^{y_x^c}
\end{equation}

where $y_x^c$ denotes the one-hot encoded Ground Truth label of the data, and $p_x^c$ denotes the predictive probability of the model for category c. By establishing a connection between the Dirichlet distribution and the confidence distribution within the evidence theory, the probabilities of different categories and the uncertainties can be derived based on the evidence collected from the network. The aforementioned equation can be further refined as follows:
\begin{align}
    \mathcal{L}_{\mathrm{ice}}\left(b_x^c, \alpha_x^c, y_x^c\right) 
    &=\!\! -\!\log \!\! \left( \!\! \int \!\! \prod_{c=1}^C \!\! \left(p_x^c\right)^{y_x^c} \!\! \frac{1}{B\left(\alpha_i\right)} \!\! \times \!\! \prod_{c=1}^C \!\! \left(p_x^c\right)^{\alpha_x^c - 1} \!\! \right) \!dp_x^c \notag \\ 
     &=\sum_{c=1}^{C}{y_x^c\left(\log{\left(S\right)}-\log{\left(\alpha_x^c\right)}\right)}
\end{align}

In order to ensure that the incorrect class produces less evidence, a KL Divergence Regularization Loss is introduced on top of this. The formulation of this loss is as follows:

\begin{align}
    \mathcal{L}_{KL}(\boldsymbol{\alpha}_x^c) = &\log \left( \frac{\Gamma \left( \sum_{c=1}^{C} \tilde{\alpha}_{i,c} \right)}{\Gamma(C) \prod_{c=1}^{C} \Gamma(\tilde{\alpha}_{i,c})} \right)
    \notag \\
    &+ \sum_{c=1}^{C} (\tilde{\alpha}_x^c - 1) \left[ \psi(\tilde{\alpha}_x^c) - \psi \left( \sum_{c=1}^{C} \tilde{\alpha}_x^c \right) \right]
\end{align}

Where $\Gamma(\cdot)$ is the Gamma function, and $\tilde{\alpha}_x^c = y_x^c + (1 - y_x^c) \odot \alpha_x^c$ denotes the adjusted parameter. This is used to ensure that the evidence for the Ground Truth does not incorrectly turn out to be 0. However, when dealing with ill-defined margins, traditional regularization losses tend to indiscriminately suppress the evidence values of incorrect classes, which in turn leads to distortions in the attention mechanisms guided by such evidence. In medical image segmentation tasks, these ill-defined margins typically correspond to pixels in the blurred boundary regions of the images, which are precisely the critical areas that require precise segmentation and attention guidance. To address this issue, we propose a novel regularization term: 

\begin{equation}
    \mathcal{L}_u(u_x, p_x^{gt}) =  (1 - p_x^{gt}) \log(u_x)
\end{equation}

This regularization term reaches its minimum value when both the uncertainty and the predicted probability tend toward 1. This design ensures that when the model encounters ill-defined boundaries, it no longer simply suppresses the evidence for incorrect classes but can also optimize by increasing predictive uncertainty. The resulting uncertainty map can provide more rational guidance for segmentation.

Therefore, it can be concluded that the final loss function can be expressed as follows:

\begin{equation}
    \mathcal{L}_{loss} = \mathcal{L}_{ice} + \lambda_1 \mathcal{L}_{KL} + \lambda_2 \mathcal{L}_u
\end{equation}

\subsubsection{Semantic-Smooth Evidence Generator}
To further enhance the network's capability for uncertainty estimation and better support the Progressive Evidential Uncertainty Guided attention mechanism, we propose a novel evidence generation function:

\begin{equation}
    Evi(\bm{x})=e^{-Relu(\bm{x})}+Relu(\bm{x})-1
\end{equation}

Specifically, as the input $x$ increases, the gradient of this function gradually rises and eventually converges to 1. This design enables the model to slow down the reduction of evidence for incorrect classes when dealing with class-confusing samples, thereby providing more authentic and reliable guidance in the attention mechanism.
\begin{table*}[htbp]
\centering 
\scalebox{0.67}{
\begin{tabular}{lrrrrrrrrrrrrrrrr}
\toprule
\multirow{2}{*}{\textbf{Model}} & \multicolumn{4}{c}{\textbf{CVC-ClinicDB}} & \multicolumn{4}{c}{\textbf{ETIS-LARIBPOLYPDB}} & \multicolumn{4}{c}{\textbf{Kvasir-SEG}} & \multicolumn{4}{c}{\textbf{ISIC2018}} \\
\cmidrule(r){2-5} \cmidrule(lr){6-9} \cmidrule(lr){10-13} \cmidrule(l){14-17} 
 & \textbf{Dice}\bm{$\uparrow$} & \textbf{IoU}\bm{$\uparrow$} & \textbf{ASSD}\bm{$\downarrow$} & \textbf{UEO}\bm{$\uparrow$}
 & \textbf{Dice}\bm{$\uparrow$} & \textbf{IoU}\bm{$\uparrow$} & \textbf{ASSD}\bm{$\downarrow$} & \textbf{UEO}\bm{$\uparrow$}
 & \textbf{Dice}\bm{$\uparrow$} & \textbf{IoU}\bm{$\uparrow$} & \textbf{ASSD}\bm{$\downarrow$} & \textbf{UEO}\bm{$\uparrow$}
 & \textbf{Dice}\bm{$\uparrow$} & \textbf{IoU}\bm{$\uparrow$} & \textbf{ASSD}\bm{$\downarrow$} & \textbf{UEO}\bm{$\uparrow$} \\
  
 \midrule
UNet[2015]& 0.8852 & 0.8082 & 1.7659 & 0.2219 & 0.6382 & 0.5477 & 12.5371 & 0.0445 & 0.7989 & 0.7033 & 2.6120 & 0.2522 & 0.9038 & 0.8353 & 0.5789 & 0.1912 \\ 
UNet++[2018]& 0.9179 & 0.8577 & 1.2173 & 0.1885 & \bluetext{0.7155} & \bluetext{0.6273} & 9.0776 & 0.0683 & 0.8499 & 0.7751 & 2.0753 & 0.2278 & \bluetext{0.9126} & \bluetext{0.8485} & \bluetext{0.3464} & 0.3333 \\ 
AttentionUNet[2018] & 0.8718 & 0.7894 & 1.5996 & 0.1989 & 0.4978 & 0.4117 & 17.285 & 0.0516 & 0.8457 & 0.7666 & \bluetext{1.9611} & 0.2305 & 0.8806 & 0.8034 & 0.7559 & \bluetext{0.3657} \\ 
SwinUNETR[2018] & 0.9119 & 0.8514 & 2.0253 & 0.0818 & 0.6366 & 0.5452 & 13.6038 & 0.0422 & 0.8086 & 0.7183 & 2.5677 & 0.2360 & 0.9015 & 0.8308 & 0.4090 & 0.3228 \\ 
EMCAD[2024] & 0.8647 & 0.7810 & 1.1005 & 0.2151 & 0.6110 & 0.5203 & \bluetext{6.6214} & \redtext{0.1705} & 0.8115 & 0.7225 & 2.4552 & \bluetext{0.2689} & 0.8754 & 0.7918 & 0.7723 & 0.3276 \\ 
U-KAN[2024] & \bluetext{0.9265} & \bluetext{0.8729} & \bluetext{0.7491} & \bluetext{0.2228} & 0.6707 & 0.5804 & 7.6526 & 0.0548 & \bluetext{0.8604} & \bluetext{0.7880} & 2.1540 & 0.2293 & 0.9109 & 0.8450 & 0.3818 & 0.3358 \\ 
\midrule
Evi U-KAN[ours] & \redtext{0.9395} & \redtext{0.8891} & \redtext{0.4169} & \redtext{0.2323} & \redtext{0.7789} & \redtext{0.7051} & \redtext{6.4050} & \bluetext{0.1500} & \redtext{0.9008} & \redtext{0.8397} & \redtext{1.3803} & \redtext{0.2882} & \redtext{0.9156} & \redtext{0.8514} & \redtext{0.3146} & \redtext{0.3872} \\ 
\bottomrule
\end{tabular}}
\caption{The segmentation results of different models on four mainstream datasets, where red represents the optimal value and blue represents the suboptimal value}
\label{tab:performance_comparison}
\end{table*}
\section{Experiments}
\subsection{Experimental Details}
We selected CVC-Clinic, ETIS-LARIBPOLYPDB, Kvasir-SEG and ISIC2018 as the experimental dataset. We selected Dice and IoU as the evaluation metrics of model segmentation accuracy, Average symmetric surface distance (ASSD) as the evaluation metrics of model segmentation accuracy at target edges, and Uncertainty-error overlap (UEO) was selected as an evaluation metric for model uncertainty estimation capability. We selected U-KAN\cite{li-etal:ukan}, EMCAD\cite{rahman-etal:emcad}, SwinUNETR\cite{hatamizadeh-etal:swin-unetr}, AttentionUNet\cite{oktay-etal:attention-unet}, UNet++\cite{zhou-etal:unet++}, and UNet\cite{ronneberger-etal:u-net} models as comparison methods. We selected Bayesian+Monte-Carlo Dropout, and Evidential Deep Learning as the comparison uncertainty estimation methods. To verify that our proposed method has better uncertainty estimation ability. All models are trained using the Adam optimizer with a learning rate of $1e-4$ and all models are trained using the evidence generator and loss function proposed in this paper, with $\lambda_1=\min{\left(1,\frac{epoch\times10}{all\ epoch}\right)}$, $\lambda_2=0.5$ in the loss function. where all epochs denotes the training the total number of epochs and epoch denotes the current number of trained epochs. All the experiments use five-fold cross-validation and take the average of the final results as the final results. In order to speed up the training, the input image is resized into 3×256×256 size and fed into the model for training.
\subsection{Evaluation Metrics}
Firstly, we selected Dice and IoU as the metrics for evaluating the accuracy of model segmentation, and they can be calculated by the following equation:

\begin{equation}
    Dice = \frac{2|R \cap G|}{|R| + |G|}
\end{equation}

\begin{equation}
    IoU = \frac{|R \cap G|}{|R \cup G|}
\end{equation}

Where R denotes the prediction of the model and G denotes Ground Truth. a larger value of Dice and IoU denotes that the model has better segmentation accuracy. \par
In addition, we still use Average Symmetric Surface Distance (ASSD) to evaluate the segmentation accuracy of the model at the edges, where $S_R$ denotes the prediction result of the model and $S_G$ is the Ground Truth, which is calculated as follows:

\begin{tiny}
\begin{equation}
    ASSD = \frac{1}{S_R + S_G} \times \left( \sum_{R \in S_R} d(R, S_G) + \sum_{G \in S_G} d(G, S_R) \right)
\end{equation}
\end{tiny}

where $d(r, S_G) = \min_{g \in S_G} (\| r - g \|) S_G$ denotes the minimum Euclidean distance from point r to all points in $S_G$. It can be posited that a smaller value of ASSD is indicative of a better model segmentation on the edges.
Finally, Uncertainty-error overlap (UEO) is selected as an indicator to evaluate the model uncertainty estimation ability, which can be calculated using the following equation:

\begin{equation}
    UEO = \frac{2|R_{\text{loss}} \cap U|}{|R_{\text{loss}}| + |U|}
\end{equation}

Where $R_{loss}$ denotes the prediction error between the model's predictions and Ground Truth, U denotes the Uncertainty Map given by the model, and a larger UEO value indicates a greater ability of the model to estimate uncertainty.

\subsection{Results}
As demonstrated in Table \ref{tab:performance_comparison}, this section presents the results of the segmentation performance and uncertainty estimation under multiple datasets by comparing the Evidential U-KAN model with the current mainstream medical image segmentation models (including the improved model based on the U-KAN architecture). The results demonstrate that, while the Evidential U-KAN model achieves significant improvement in both segmentation accuracy (measured by Dice and IoU) and edge segmentation performance (measured by ASSD), it also exhibits better uncertainty estimation capability.\par
Specifically, the Evidential U-KAN model outperforms the U-KAN architecture in terms of average ASSD value on several mainstream medical image segmentation datasets, while its Dice improves by an average of 4.16\% and IoU improves by an average of 4.98\% compared to the baseline method. Furthermore, the Evidential U-KAN model demonstrates a 5.38\% enhancement in the uncertainty evaluation metric, providing additional validation of its superiority in the trusted segmentation task. This finding underscores the efficacy of our methods, which significantly enhance the model's capacity for learning in challenging regions and its plausible deployment capability.

\begin{figure}
    \centering
    \includegraphics[width=1\linewidth, angle=0]{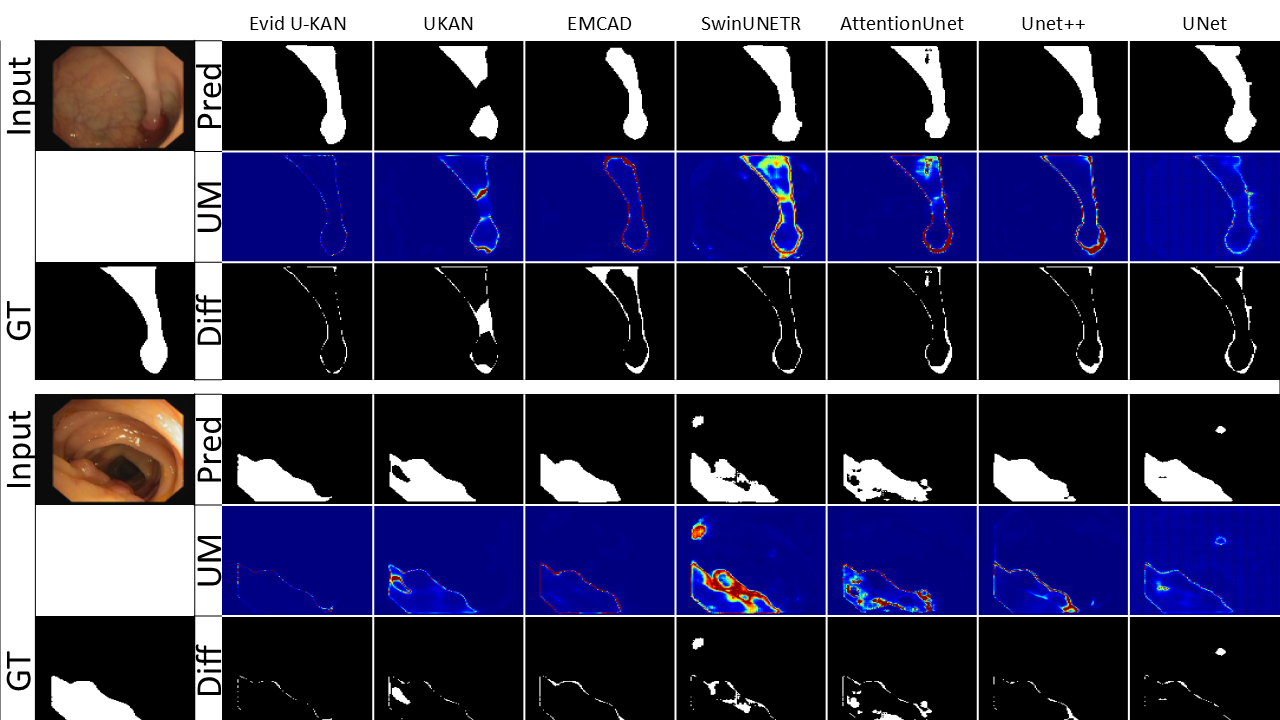}
    \caption{Visualization display of segmentation results of different models on the CVC-linicDB dataset, where Input represents the input image, GT represents the Ground truth, Pred represents the model segmentation result, UM represents the Uncertainty Map, and Diff represents the error between the segmentation result and the Ground truth}
    \label{fig:result}
\end{figure}

As demonstrated in Figure \ref{fig:result}, the visualization of the segmentation results of different models on the CVC-ClinicDB dataset, including the segmentation output, the Uncertainty Map and the error (Loss) between the prediction results and Ground Truth. Through comparison, it is intuitively found that the Evidential U-KAN model significantly outperforms other models in segmentation at the edges of the lesion region, and its segmentation results are closer to Ground Truth, especially in the prediction of fine boundaries, which demonstrates a significant advantage. \par
Furthermore, all models demonstrated a high degree of similarity between the Uncertainty Map and the prediction error. The findings of this analysis suggest that this phenomenon is primarily attributable to the incorporation of a regularization constraint term within the loss function, which serves to augment the model's capacity to capture and articulate uncertainty information. This mechanism enhances the robustness of the model to the segmentation task and enables it to exhibit higher confidence when dealing with complex boundaries and difficult-to-segment regions.\par
 
\begin{table}[htbp]
\centering
\scalebox{0.65}{ 
    \begin{tabular}{rrrrrr}
    \toprule
    \multirow{2}{*}{\textbf{EUGA}} & \multirow{2}{*}{\textbf{SAEL}} & \multicolumn{4}{c}{\textbf{CVC-ClinicDB}} \\
     \cmidrule(lr){3-6}
     & & \textbf{Dice}\bm{$\uparrow$} & \textbf{IoU}\bm{$\uparrow$} & \textbf{ASSD}\bm{$\downarrow$} & \textbf{UEO}\bm{$\uparrow$}\\
    \midrule
     & & 0.9347 & 0.8823 & 0.5257 & 0.1887 \\
    \checkmark & & \redtext{0.9407} & \redtext{0.8914} & \redtext{0.3540} & 0.1699 \\
     & \checkmark & 0.9275 & 0.8739 & 1.4697 & \bluetext{0.2179}\\
    \checkmark & \checkmark & \bluetext{0.9395} & \bluetext{0.8891} & \bluetext{0.4169} & \redtext{0.2323} \\
    \bottomrule
    \end{tabular}
} 
\caption{Our model performs ablation experiments on the CVC LinicDB dataset, where red represents the optimal value and blue represents the suboptimal value}
\label{tab:ablation}
\end{table}

On this basis, we also conducted ablation experiments on the CVC-Clinic dataset, and conducted ablation experiments on the method proposed in this paper. Ablation experiments are conducted to verify the effectiveness of each module. As demonstrated in Table \ref{tab:ablation}, the proposed EUGA module has been shown to have a significant impact on enhancing the segmentation accuracy of the model. By introducing the Uncertainty Map to guide the segmentation process, the EUGA module enhances the model's capacity to focus on difficult regions, especially the edges.\par
Moreover, the findings presented in the table demonstrate that the incorporation of SAEL results in a modest decline in segmentation accuracy; however, this trade-off considerably augments the model's capacity to discern uncertainty. This enhancement enables the model to more accurately identify regions of uncertainty in the prediction results, thereby enhancing the credibility and reliability of the model in  real-world clinical applications.\par
\begin{table}[htbp]
\centering
\scalebox{0.58}{
\begin{tabular}{llrrrrrrrr}
\toprule
\multirow{2}{*}{\textbf{Method}} & \multirow{2}{*}{\textbf{Model}} & \multicolumn{4}{c}{\textbf{Original}} & \multicolumn{4}{c}{\textbf{Noise}} \\
\cmidrule(lr){3-6} \cmidrule(lr){7-10}
& & \textbf{Dice}\bm{$\uparrow$} & \textbf{IoU}\bm{$\uparrow$} & \textbf{ASSD}\bm{$\downarrow$} & \textbf{UEO}\bm{$\uparrow$} & \textbf{Dice}\bm{$\uparrow$} & \textbf{IoU}\bm{$\uparrow$} & \textbf{ASSD}\bm{$\downarrow$} & \textbf{UEO}\bm{$\uparrow$} \\
\midrule
\multirow{2}{*}{Bayes} & UNet & 0.8945 & 0.8200 & 1.6342 & 0.0253 & 0.3673 & 0.3138 & 12.8478 & 0.1238 \\
& U-KAN& 0.9271 & 0.8734 & 1.0555 & 0.0263 & 0.3997 & 0.3234 & 14.1788 & 0.1107 \\
\midrule
\multirow{3}{*}{EDL} & UNet & 0.8817 & 0.8043 & 1.8939 & 0.1650 & 0.3998 & 0.2964 & 9.9037 & 0.0882 \\
& U-KAN& 0.9298 & 0.8757 & 0.5845 & 0.1966 & 0.4315 & 0.3470 & 9.6523 & 0.1369 \\
& Evi U-KAN & \redtext{0.9407} & \redtext{0.8914} & \redtext{0.3540} & 0.2002 & \bluetext{0.5720} & \bluetext{0.4782} & \bluetext{6.0838} & 0.1428 \\
\midrule
\multirow{3}{*}{SAEL}& UNet & 0.8852 & 0.8082 & 1.7659 & 0.2219 & 0.3882 & 0.2837 & 11.3078 & 0.0801 \\
& U-KAN& 0.9265 & 0.8729 & 0.7491 & \bluetext{0.2228} & 0.4374 & 0.3600 & 7.4548 & \bluetext{0.1649} \\
& Evi U-KAN & \bluetext{0.9395} & \bluetext{0.8891} & \bluetext{0.4169} & \redtext{0.2323} & \redtext{0.6281} & \redtext{0.5277} & \redtext{5.2947} & \redtext{0.1681} \\
\bottomrule
\end{tabular}
}
\caption{The results of training different models on the CVC ClinicDB dataset, where Original is the original test set of CVC ClinicDB, Noise is the result of adding random Gaussian noise with a mean of 0 and a standard deviation of 0.1 to 0.4 during testing on the test set, red represents the optimal value, and blue represents the suboptimal value.}
\label{tab:noise}
\end{table}
\begin{table}[htbp]
 \centering
 \scalebox{0.55}{
 \begin{tabular}{lrrrrrrrr}
  \toprule
  \textbf{Train dataset}& \multicolumn{4}{c}{\textbf{CVC-ClinicDB}}& \multicolumn{4}{c}{\textbf{Kvasir-SEG}}\\
  \midrule
 \textbf{Test dataset}& \multicolumn{4}{c}{\textbf{Kvasir-SEG}} & \multicolumn{4}{c}{\textbf{CVC-ClinicDB}} \\
 \cmidrule(lr){2-5} \cmidrule(lr){6-9}
& \textbf{Dice}\bm{$\uparrow$} & \textbf{IoU}\bm{$\uparrow$} & \textbf{ASSD}\bm{$\downarrow$} & \textbf{UEO}\bm{$\uparrow$} & \textbf{Dice}\bm{$\uparrow$} & \textbf{IoU}\bm{$\uparrow$} & \textbf{ASSD}\bm{$\downarrow$} & \textbf{UEO}\bm{$\uparrow$}  \\
 \midrule
 UNet[2015] & 0.4638& 0.3382 & 10.7495 & 0.0798 & 0.4705& 0.3406& 10.5420& 0.0881 \\
 UNet++[2018] & 0.6011& 0.4871& 7.2193& 0.1939& \bluetext{0.6479}& \bluetext{0.5476}& \redtext{5.3155} & 0.2274 \\
 AttentionUNet[2018] & \bluetext{0.6449} & \bluetext{0.5263} & \redtext{5.0402} & \redtext{0.3469} & 0.5299& 0.4274& 8.8125 & 0.1126 \\
 SwinUNETR[2022]& 0.4416 & 0.3169 & 11.5255 & 0.1668 & 0.5124& 0.3946& 7.7758& \bluetext{0.2863}\\
 EMCAD[2024]& 0.4896 & 0.3738 & 7.0906 & 0.1714 & 0.4801& 0.3701& 10.8616 & 0.1377 \\
 U-KAN[2024] & 0.6398 & 0.5244 & 7.6376 & 0.2883 & 0.6274 & 0.5266& 7.0000& 0.2810 \\
 Evi U-KAN[ours] & \redtext{0.6781} & \redtext{0.5701} & \bluetext{5.7221} & \bluetext{0.3233} & \redtext{0.6802}& \redtext{0.5727}& \bluetext{6.5950}& \redtext{0.3390} \\
 \bottomrule
 \end{tabular}}
  \caption{The segmentation performance of different models across datasets, where red represents the optimal value and blue represents the suboptimal value}
  \label{tab:crsoo_dataset}
 \end{table}

In addition, the model was compared with several uncertainty estimation methods, and the CVC-Clinic dataset was selected for experimental evaluation. The results of these experiments are presented in Table \ref{tab:noise}. It can be seen that the semantic-preserving evidence learning strategy proposed in this paper have been shown to significantly enhance the model’s capability of uncertainty estimation. The optimal design of the evidence generator enables the model to more accurately characterize the uncertainty region of the prediction results. Concurrently, the introduction of the evidence regularization term further enhances the accuracy and stability of the uncertainty estimation by effectively fusing the a priori information and reasonably constraining the evidence distribution of the model. In terms of segmentation accuracy, the experimental results show that our method not only exhibits higher accuracy in the segmentation of edge details, but also outperforms the existing mainstream methods in terms of overall segmentation performance.\par
As demonstrated in Table \ref{tab:crsoo_dataset}, the experimental results show the excellent generalization capability of Evidential U-KAN across different datasets. When trained on the CVC-ClinicDB dataset and tested on Kvasir-SEG, Evi U-KAN achieves a Dice score of 0.6781, which is significantly higher than the scores achieved by other models, including UNet (0.4638), UNet++ (0.6011), and Attention UNet (0.6449). A similar outcome is observed when the model is trained on Kvasir-SEG and tested on CVC-ClinicDB, attaining a Dice score of 0.6802. This result once again surpasses the performance of existing models, including UNet (0.4705), UNet++ (0.6479), and Attention UNet (0.5299). These results indicate that Evi U-KAN not only excels in domain-specific tasks, but also generalizes well to unseen data, highlighting its robustness and adaptability in diverse medical image segmentation scenarios.

\section{Conclusions}
In this paper, we propose a trustworthy medical image segmentation method based on progressive evidential uncertainty guidance, which innovatively combines the Progressive Evidence Uncertainty guided Attention mechanism and the semantic-preserving evidence learning strategy to construct a new segmentation framework. This method, while achieving more accurate uncertainty estimation, makes effective use of the Uncertainty Map, significantly enhancing the model's ability to focus on difficult regions. Experimental results on mainstream datasets including CVC-ClinicDB, and Kvasir-SEG show that the method we proposed in this paper achieves excellent performance in terms of segmentation accuracy and uncertainty estimation capability. We acknowledge that, due to its iterative nature, this method incurs a higher computational cost in comparison with alternative approaches and is unable to distinguish unlearnable regions within difficult regions. This will be a focus of our future efforts.

\bibliographystyle{named}
\bibliography{refs}

\begin{thebibliography}{}

\bibitem[\protect\citeauthoryear{Amersfoort \bgroup \em et al.\egroup }{2020}]{vanamersfoort-etal:uncertainty-estimation}
J.~Van Amersfoort, L.~Smith, Y.~W. Teh, et~al.
\newblock Uncertainty estimation using a single deep deterministic neural network.
\newblock In {\em International Conference on Machine Learning}, pages 9690--9700. PMLR, 2020.

\bibitem[\protect\citeauthoryear{Bao \bgroup \em et al.\egroup }{2021}]{bao-etal:evidential-open-set}
W.~Bao, Q.~Yu, and Y.~Kong.
\newblock Evidential deep learning for open set action recognition.
\newblock In {\em Proceedings of the IEEE/CVF International Conference on Computer Vision}, pages 13349--13358, 2021.

\bibitem[\protect\citeauthoryear{Cao \bgroup \em et al.\egroup }{2022}]{cao-etal:swin-unet}
H.~Cao, Y.~Wang, J.~Chen, et~al.
\newblock Swin-unet: Unet-like pure transformer for medical image segmentation.
\newblock In {\em European Conference on Computer Vision}, pages 205--218. Springer Nature Switzerland, 2022.

\bibitem[\protect\citeauthoryear{Chen \bgroup \em et al.\egroup }{2024a}]{chen-etal:transunet}
J.~Chen, J.~Mei, X.~Li, et~al.
\newblock Transunet: Rethinking the u-net architecture design for medical image segmentation through the lens of transformers.
\newblock {\em Medical Image Analysis}, 97:103280, 2024.

\bibitem[\protect\citeauthoryear{Chen \bgroup \em et al.\egroup }{2024b}]{chen-etal:uncertainty-aware}
Y.~Chen, Z.~Yang, C.~Shen, et~al.
\newblock Evidence-based uncertainty-aware semi-supervised medical image segmentation.
\newblock {\em Computers in Biology and Medicine}, 170:108004, 2024.

\bibitem[\protect\citeauthoryear{Christ and Parvathi}{2011}]{christ2011segmentation}
MC~Jobin Christ and RMS Parvathi.
\newblock Segmentation of medical image using clustering and watershed algorithms.
\newblock {\em American Journal of Applied Sciences}, 8(12):1349, 2011.

\bibitem[\protect\citeauthoryear{Dosovitskiy \bgroup \em et al.\egroup }{2020}]{dosovitskiy-etal:image-worth}
A.~Dosovitskiy, L.~Beyer, A.~Kolesnikov, et~al.
\newblock An image is worth 16x16 words: Transformers for image recognition at scale.
\newblock In {\em International Conference on Learning Representations}, 2020.

\bibitem[\protect\citeauthoryear{Han \bgroup \em et al.\egroup }{2022}]{han-etal:multi-view}
Z.~Han, C.~Zhang, H.~Fu, et~al.
\newblock Trusted multi-view classification with dynamic evidential fusion.
\newblock {\em IEEE Transactions on Pattern Analysis and Machine Intelligence}, 45(2):2551--2566, 2022.

\bibitem[\protect\citeauthoryear{Hatamizadeh \bgroup \em et al.\egroup }{2021}]{hatamizadeh-etal:swin-unetr}
A.~Hatamizadeh, V.~Nath, Y.~Tang, et~al.
\newblock Swin unetr: Swin transformers for semantic segmentation of brain tumors in mri images.
\newblock In {\em International MICCAI BrainLesion Workshop}, pages 272--284. Springer International Publishing, 2021.

\bibitem[\protect\citeauthoryear{Huang \bgroup \em et al.\egroup }{2020}]{huang-etal:unet3}
H.~Huang, L.~Lin, R.~Tong, et~al.
\newblock Unet 3+: A full-scale connected unet for medical image segmentation.
\newblock In {\em ICASSP 2020-2020 IEEE International Conference on Acoustics, Speech and Signal Processing (ICASSP)}, pages 1055--1059. IEEE, 2020.

\bibitem[\protect\citeauthoryear{Isensee \bgroup \em et al.\egroup }{2018}]{isensee-etal:nnu-net}
F.~Isensee, J.~Petersen, A.~Klein, et~al.
\newblock nnu-net: Self-adapting framework for u-net-based medical image segmentation.
\newblock {\em arXiv preprint}, arxiv:1809.10486, 2018.

\bibitem[\protect\citeauthoryear{Kendall and Gal}{2017}]{kendall-gal:bayesian-uncertainties}
A.~Kendall and Y.~Gal.
\newblock What uncertainties do we need in bayesian deep learning for computer vision?
\newblock {\em Advances in Neural Information Processing Systems}, 30, 2017.

\bibitem[\protect\citeauthoryear{Lakshminarayanan \bgroup \em et al.\egroup }{2017}]{lakshminarayanan-etal:deep-ensembles}
B.~Lakshminarayanan, A.~Pritzel, and C.~Blundell.
\newblock Simple and scalable predictive uncertainty estimation using deep ensembles.
\newblock {\em Advances in Neural Information Processing Systems}, 30, 2017.

\bibitem[\protect\citeauthoryear{Li \bgroup \em et al.\egroup }{2024}]{li-etal:ukan}
C.~Li, X.~Liu, W.~Li, et~al.
\newblock U-kan makes strong backbone for medical image segmentation and generation.
\newblock {\em arXiv preprint}, arxiv:2406.02918, 2024.

\bibitem[\protect\citeauthoryear{Liu \bgroup \em et al.\egroup }{2021}]{liu-etal:satellite-marsh}
M.~Liu, B.~Fu, E.~S, et~al.
\newblock Comparison of multi-source satellite images for classifying marsh vegetation using deeplabv3 plus deep learning algorithm.
\newblock {\em Ecological Indicators}, 125:107562, 2021.

\bibitem[\protect\citeauthoryear{Liu \bgroup \em et al.\egroup }{2024}]{liu-etal:kan}
Z.~Liu, Y.~Wang, S.~Vaidya, et~al.
\newblock Kan: Kolmogorov-arnold networks.
\newblock {\em arXiv preprint}, arxiv:2404.19756, 2024.

\bibitem[\protect\citeauthoryear{Lou \bgroup \em et al.\egroup }{2021}]{lou-etal:dc-unet}
A.~Lou, S.~Guan, and M.~Loew.
\newblock Dc-unet: Rethinking the u-net architecture with dual channel efficient cnn for medical image segmentation.
\newblock In {\em Medical Imaging 2021: Image Processing}, volume 11596, pages 758--768. SPIE, 2021.

\bibitem[\protect\citeauthoryear{Oktay \bgroup \em et al.\egroup }{2018}]{oktay-etal:attention-unet}
O.~Oktay, J.~Schlemper, L.~L. Folgoc, et~al.
\newblock Attention u-net: Learning where to look for the pancreas.
\newblock {\em arXiv preprint}, arxiv:1804.03999, 2018.

\bibitem[\protect\citeauthoryear{Rahman \bgroup \em et al.\egroup }{2024}]{rahman-etal:emcad}
M.~M. Rahman, M.~Munir, and R.~Marculescu.
\newblock Emcad: Efficient multi-scale convolutional attention decoding for medical image segmentation.
\newblock In {\em Proceedings of the IEEE/CVF Conference on Computer Vision and Pattern Recognition}, pages 11769--11779, 2024.

\bibitem[\protect\citeauthoryear{Ronneberger \bgroup \em et al.\egroup }{2015}]{ronneberger-etal:u-net}
O.~Ronneberger, P.~Fischer, and T.~Brox.
\newblock U-net: Convolutional networks for biomedical image segmentation.
\newblock In {\em Medical Image Computing and Computer-Assisted Intervention–MICCAI 2015: 18th International Conference}, pages 234--241, Munich, Germany, October 2015. Springer International Publishing.

\bibitem[\protect\citeauthoryear{Salman \bgroup \em et al.\egroup }{2015}]{salman2015medical}
NH~Salman, BM~Ghafour, and Gullanar~M Hadi.
\newblock Medical image segmentation based on edge detection techniques.
\newblock {\em Advances in Image and Video Processing}, 3(2):1--9, 2015.

\bibitem[\protect\citeauthoryear{Sensoy \bgroup \em et al.\egroup }{2018}]{sensoy-etal:evidential}
M.~Sensoy, L.~Kaplan, and M.~Kandemir.
\newblock Evidential deep learning to quantify classification uncertainty.
\newblock {\em Advances in Neural Information Processing Systems}, 31, 2018.

\bibitem[\protect\citeauthoryear{Vaswani}{2017}]{vaswani:attention}
A.~Vaswani.
\newblock Attention is all you need.
\newblock {\em Advances in Neural Information Processing Systems}, 2017.

\bibitem[\protect\citeauthoryear{Yang \bgroup \em et al.\egroup }{2023}]{yang-etal:lung-nodule}
H.~Yang, Q.~Wang, Y.~Zhang, et~al.
\newblock Lung nodule segmentation and uncertain region prediction with an uncertainty-aware attention mechanism.
\newblock {\em IEEE Transactions on Medical Imaging}, 43(4):1284--1295, 2023.

\bibitem[\protect\citeauthoryear{Yurtkulu \bgroup \em et al.\egroup }{2019}]{yurtkulu-etal:deepLabv3}
S.~C. Yurtkulu, Y.~H. Şahin, and G.~Unal.
\newblock Semantic segmentation with extended deeplabv3 architecture.
\newblock In {\em 2019 27th Signal Processing and Communications Applications Conference (SIU)}, pages 1--4. IEEE, 2019.

\bibitem[\protect\citeauthoryear{Zhang \bgroup \em et al.\egroup }{2015}]{zhang2015medical}
Xiaoli Zhang, Xiongfei Li, and Yuncong Feng.
\newblock A medical image segmentation algorithm based on bi-directional region growing.
\newblock {\em Optik}, 126(20):2398--2404, 2015.

\bibitem[\protect\citeauthoryear{Zhou \bgroup \em et al.\egroup }{2018}]{zhou-etal:unet++}
Z.~Zhou, M.~M.~Rahman Siddiquee, N.~Tajbakhsh, et~al.
\newblock Unet++: A nested u-net architecture for medical image segmentation.
\newblock In {\em Deep Learning in Medical Image Analysis and Multimodal Learning for Clinical Decision Support: 4th International Workshop, DLMIA 2018, and 8th International Workshop, ML-CDS 2018, Held in Conjunction with MICCAI 2018}, pages 3--11, Granada, Spain, September 2018. Springer International Publishing.

\bibitem[\protect\citeauthoryear{Zhou \bgroup \em et al.\egroup }{2021}]{zhou-etal:nnformer}
H.~Y. Zhou, J.~Guo, Y.~Zhang, et~al.
\newblock nnformer: Interleaved transformer for volumetric segmentation.
\newblock {\em arXiv preprint}, arxiv:2109.03201, 2021.

\bibitem[\protect\citeauthoryear{Zou \bgroup \em et al.\egroup }{2022}]{zou-etal:tbrats}
K.~Zou, X.~Yuan, X.~Shen, et~al.
\newblock Tbrats: Trusted brain tumor segmentation.
\newblock In {\em International Conference on Medical Image Computing and Computer-Assisted Intervention}, pages 503--513. Springer Nature Switzerland, 2022.

\end{thebibliography}

\end{document}